\documentclass[conference]{IEEEtran}
\IEEEoverridecommandlockouts

\usepackage{cite}
\usepackage{amsmath,amssymb,amsfonts}
\usepackage{algorithmic}
\usepackage{graphicx}
\usepackage{textcomp}
\usepackage{xcolor}

\usepackage{multirow}
\usepackage{subfigure}
\usepackage{graphicx}
\usepackage{makecell}
\usepackage{hyperref}

\DeclareRobustCommand*{\IEEEauthorrefmark}[1]{%
    \raisebox{0pt}[0pt][0pt]{\textsuperscript{\footnotesize\ensuremath{#1}}}}

\def\BibTeX{{\rm B\kern-.05em{\sc i\kern-.025em b}\kern-.08em
    T\kern-.1667em\lower.7ex\hbox{E}\kern-.125emX}}
\begin{document}

\title{BP-GPT: Auditory Neural Decoding Using fMRI-prompted LLM}

\author{\IEEEauthorblockN{Xiaoyu Chen\textsuperscript{1, 2} \qquad Changde Du\textsuperscript{1, 2} \qquad Che Liu\textsuperscript{1, 3} \qquad Yizhe Wang\textsuperscript{4} \qquad Huiguang He{$^*$}\textsuperscript{1, 2}}

\IEEEauthorblockA{\textit{\IEEEauthorrefmark{1}  NeuBCI Group, State Key Laboratory of Brain Cognition and Brain-inspired Intelligence Technology, CASIA} \\
\textit{\IEEEauthorrefmark{2}  School
of Artificial Intelligence, University of Chinese Academy of Sciences}\\
\textit{\IEEEauthorrefmark{3}  School of Future Technology, University of Chinese Academy of Sciences}\\
\textit{\IEEEauthorrefmark{4} State Key
Laboratory of Multimodal Artificial
Intelligence Systems, CASIA}
} 
}

\maketitle

\begin{abstract}
Decoding language information from brain signals represents a vital research area within brain-computer interfaces, particularly in the context of deciphering the semantic information from the fMRI signal. Although existing work uses LLM to achieve this goal, their method does not use an end-to-end approach and avoids the LLM in the mapping of fMRI-to-text, leaving space for the exploration of the LLM in auditory decoding. In this paper, we introduce a novel method, the \textbf{Brain Prompt GPT (BP-GPT)}. By using the brain representation that is extracted from the fMRI as a prompt, our method can utilize GPT-2 to decode fMRI signals into stimulus text. Further, we introduce the text prompt and align the fMRI prompt to it. By introducing the text prompt, our BP-GPT can extract a more robust brain prompt and promote the decoding of pre-trained LLM. We evaluate our BP-GPT on the open-source auditory semantic decoding dataset and achieve a significant improvement up to $\textbf{4.61\%}$ on METEOR and $\textbf{2.43\%}$ on BERTScore across all the subjects compared to the state-of-the-art method. The experimental results demonstrate that using brain representation as a prompt to further drive LLM for auditory neural decoding is feasible and effective. The code is available at https://github.com/1994cxy/BP-GPT.
\end{abstract}


\section{Introduction}
“The limits of my language mean the limits of my world” - Ludwig Wittgenstein. Wittgenstein’s statement refers to the standpoint that a person's entire understanding of the world is reflected in the things they can describe in language. So it is important for human-centric artificial intelligence, for example, the brain-computer interface, to understand the language information from the human brain. 

In recent years, due to the rapid development of deep learning and its widespread application in brain-computer interfaces, especially in neural encoding and decoding \cite{scotti2024reconstructing,lin2022mind,du2023decoding,li2022multi,wang2022multi}, significant progress has been made in the research field of extracting language information from brain signals. For example, reconstructing audio of auditory stimuli from brain signals \cite{yang2015speech,pasley2012reconstructing,santoro2017reconstructing,defossez2023decoding}, or reconstructing corresponding text \cite{affolter2020brain2word,defossez2023decoding,pereira2018toward,xi2023unicorn,tang2023semantic}. In this article, we focus on reconstructing language information by decoding the original text. Specifically, we focus on text decoding in auditory neural decoding scenarios. This experimental paradigm has been proven to have potential applications in decoding imagined speech \cite{tang2023semantic} and may help computers understand your thoughts in the future. For the types of brain-computer interfaces, we focus on decoding text from functional magnetic resonance imaging (fMRI), a widely used non-invasive brain-computer interface in both research and practical applications. 

To accomplish this goal, two primary challenges must be addressed. Firstly, the temporal resolution of fMRI signals presents a significant obstacle. Despite its commendable spatial resolution and non-invasive characteristics, fMRI signals exhibit significantly low temporal resolution. For instance, in the context of commonly spoken English, where the average speaking speed exceeds 2 words per second \cite{tang2023semantic}, the BOLD response to neural activity rises and falls over approximately 10 seconds which is extremely slower than the speech stimuli \cite{logothetis2003underpinnings}. Secondly, another essential challenge is the significant difference between fMRI modality and text modality. In fact, in auditory information decoding scenarios, the text does not present as the stimulus signal received by the subject. Rather, it contains the semantics of the stimulus signal. Therefore, the quality of modal alignment significantly influences the effectiveness of the decoding model.

\begin{figure}[t]
    \centering
    \includegraphics[width=\linewidth]{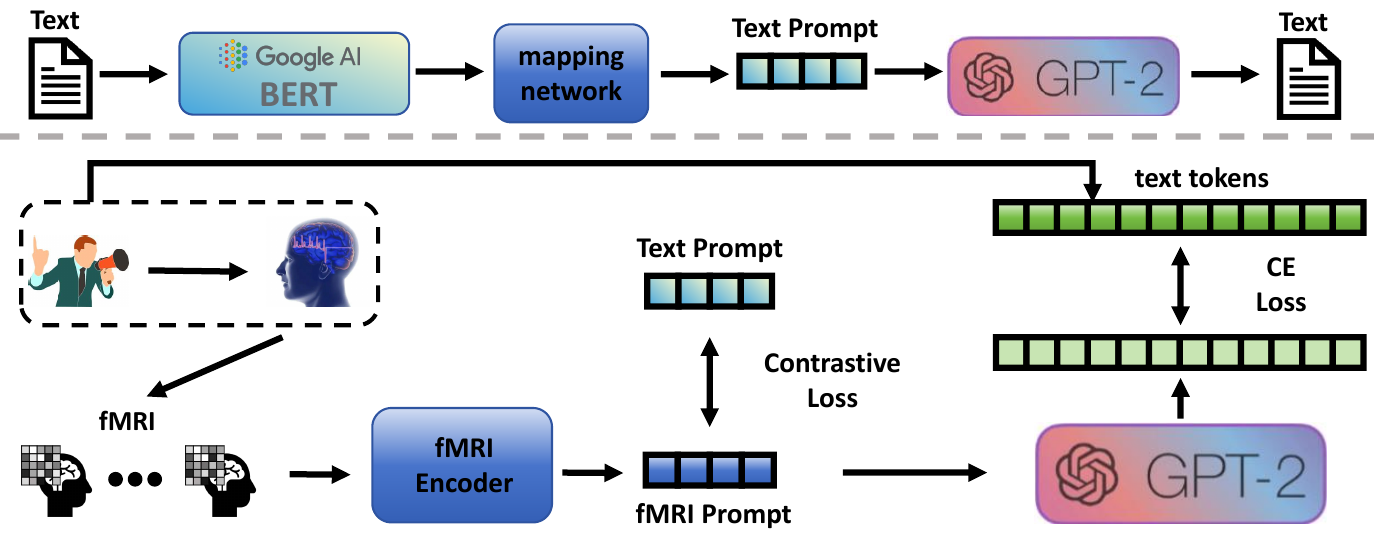}
    \caption{Two training stages of our method.}
    \label{fig: training}
\end{figure}

The low temporal resolution in fMRI necessitates our model to decode multiple words from a single fMRI repetition time (TR) during the decoding process. To tackle this ill-posed inverse problem, we propose a solution by incorporating a language prior through a pre-trained Large Language Model (LLM) - Generative Pre-trained Transformer 2 (GPT-2) \cite{radford2019language}. As illustrated in the bottom part of Fig \ref{fig: training}, we establish a mapping from fMRI to prompts and train GPT-2 to employ autoregressive methods, generating corresponding text based on the provided fMRI prompts. By applying the cross-entropy loss to the output logit of GPT-2, the fMRI encoder can learn the suitable prompts for the target text. 

To mitigate the substantial modal disparity between the fMRI signal and the text data, we introduced another text prompt during the training. As depicted in the upper part of Fig \ref{fig: training}, we divided the training into two stages. In the first stage, we trained a mapping network to extract the text prompt using the features extracted by Bidirectional Encoder Representations from Transformers (BERT) \cite{devlin2018bert}. Subsequently, we reconstructed the text using GPT-2 in a manner similar to the fMRI-to-text. Since the input and output in this stage are identical, eliminating modal differences, we designate the text prompt as the optimal prompt for the ground-truth text. In the second stage, we learn the fMRI prompt as described above and use contrastive learning to align the fMRI prompt with the text prompt, enhancing decoding performance consequently.

Compared with recent work \cite{tang2023semantic}, our model not only adopts an end-to-end approach, allowing LLM to participate in the mapping process from fMRI to text, but also introduces contrastive learning with the text prompt to further reduce the impact of modal differences on decoding performance. In summary, our main contributions are as follows: 1) We propose \textbf{Brain Prompt GPT (BP-GPT)}, which is a novel structure that can use the fMRI prompt to decode the text of speech stimuli in an end-to-end structure. 2) We introduced a language prior by a pre-trained LLM (GPT-2) to compensate for the low temporal resolution of fMRI; through contrastive learning, we encourage fMRI prompts to align with text prompts, thereby reducing the impact of modal differences on decoding performance. 3) We evaluated our BP-GPT model on an open-source auditory semantic decoding dataset and the results demonstrate the feasibility and advantages of our approach.


\section{Method}
\subsection{fMRI to Text Decoding}
In this section, we will introduce our method which could address the two challenges mentioned above. Our method consists of two essential components. The first is the fMRI-prompted text decoding method, which introduces the prompt paradigm into fMRI decoding and addresses the problem of low temporal resolution of fMRI. The second is aligning the fMRI prompt to the text prompt, thereby reducing the impact of modal differences and addressing the last challenge. 

\subsubsection{fMRI-prompted text decoding}
We use an fMRI encoder model to encode the fMRI into the representation space and use this representation as the prompt of the text generation process of GPT-2:

\begin{equation}
    \mathit{P}_{i}^B = \mathbf{E}_{\eta}(x_{i}^B),
\end{equation}
where the $\mathbf{E}_{\eta}$ is the fMRI encoder, and $x_{i}^B$ denote the fMRI. $\mathit{P}_{i}^B = (p^B_{1}, \cdots, p^B_{k})$ denote the fMRI prompt which is extracted by the fMRI encoder.

Since we did not introduce a pre-trained fMRI encoder here, the fMRI-to-text part did not require a mapping network as it was used in the extraction of text prompt. Furthermore, the fMRI encoder can be trained using the cross-entropy loss on the output of the GPT-2 with the following form:


\begin{align}\label{loss:CE}
    \mathcal{L}_{brain} &= -\sum_{i=1}^{N}\log p_{\eta}(\mathit{W}|\mathit{P}_{i}^B)\\
    &= -\sum_{i=1}^{N}\sum_{j=1}^{\mathcal{L}}\log p_{\eta}(w_{j}|p^B_{1}, \ldots, p^B_{k}, w_{1}, \ldots, w_{j-1}).
    \notag
\end{align}


\subsubsection{Align with the Optimal Prompt}
Given the considerable significant modal differences between fMRI and text, extracting effective fMRI prompts through the fMRI encoder poses challenges. Conversely, learning a text prompt in the text reconstruction task inherently circumvents this modal difference. Thus, we argue that the text prompt can be deemed as the optimal prompt for the text to be generated. To utilize the optimal prompt, we employ contrastive learning to align the fMRI prompt with the text prompt.

Specifically, we set the fMRI prompt and text prompt of the same text as the positive pair and calculate the similarity between the positive pair using the following formula:

\begin{equation}
    \mathit{S}_{p} = \exp (cos(\mathit{P}^{i}_B \cdot \mathit{P}^{i}_T)/\tau),
\end{equation}
where the $\tau$ refers to the temperature hyperparameter.

For the negative pair, we use the fMRI prompt and text prompt from different samples, and the fMRI prompt of different samples. The similarity between the negative pairs can be formulated as:

\begin{equation}
\begin{split}
    \mathit{S}_{n} = &\exp (cos(\mathit{P}^{i}_B \cdot \mathit{P}^{j}_B)/\tau) +  \exp (cos(\mathit{P}^{i}_B \cdot \mathit{P}^{j}_T)/\tau), i\neq j.
\end{split}
\end{equation}

Based on the definition above, the contrastive loss has the following form:

\begin{equation}
    L_{\mathcal{C}} = -\mathbb{E}\left[\log\frac{S_p}{S_n}\right].
\end{equation}

\subsection{Training}
We divide the training process into two stages. In the initial stage, the text prompt is learned through text encoding and decoding. The training loss for the text prompt is set in the same way as the fMRI prompt in Eq. \ref{loss:CE}. Then, we train our decoding model in the second stage using the following formula:
\begin{align}
    L = L_{brain} + \alpha L_{C},
\end{align}
where the $\alpha$ is the weight for the contrastive loss.

\begin{figure}[t]
    \centering
    \includegraphics[width=\linewidth]{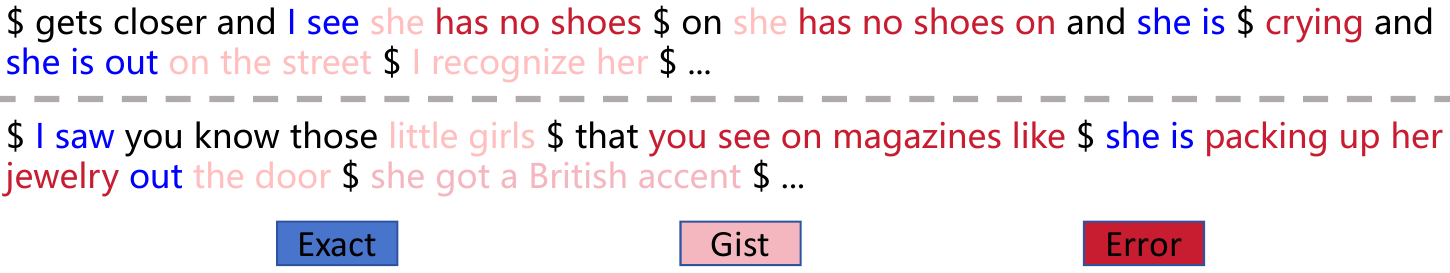}
    \caption{Ground Truth (top) and the decoding result (bottom) from the test set. \$ is added according to the fMRI TR as the alignment token.}
    \label{fig: sample}
\end{figure}

\subsection{Inference}\label{sec: inference}
As our focus is on text decoding within an auditory neural decoding scenario, the auditory stimuli received by the subjects exclusively consist of words without punctuation. This will cause trouble for the model during the inference phase, as we cannot determine the end of generation through the stop token (usually, periods are used). Although punctuation can be manually added during text annotation for audio stimuli, this introduces two concerns. Firstly, the speaker might have delivered a spontaneous speech without adhering to the ground-truth speech draft. Secondly, since the decoding often uses the fMRI signals within a fixed-length window, and the end of this window does not always match the end of the sentence, even if we add the punctuation manually, they are very likely unable to indicate the end of the generation.

The current solution adopted in the recent work \cite{tang2023semantic} is to utilize a word rate model to predict the number of words perceived by participants. The text generation process will be stopped when the length of the generated text meets the word count predicted by the word rate model. Although this approach can solve the problem, it does not fully utilize the characteristics of LLM. In our implementation (see Fig \ref{fig: sample}), we add \$ as an alignment token to the ground-truth text according to the repetition time (TR) of fMRI and stop the generation when we meet enough \$ in the generated text.

\section{Experiment}
\subsection{Dataset}
We evaluate our method on an fMRI dataset obtained during a passive natural language listening task \cite{lebel2023natural} along with its extended data \cite{tang2023semantic}. This dataset comprises fMRI data from 8 subjects recorded while they passively listened to naturally spoken English stories. The stories were sourced from \textit{The Month} and \textit{New York Times Modern Love} podcasts. Specifically, the first 3 subjects (UTS1 to UTS3) in the dataset had access to stories from both \textit{The Month} and \textit{New York Times Modern Love}, expanding the total number of stories to 84 for these three subjects. To maintain consistency with existing works \cite{lebel2023natural,tang2023semantic}, we chose the data from the first 3 subjects which are more adequate for the deep learning methods.

\subsection{Implementing Details}
The temperature of the contrastive loss is set to $\tau=0.1$, and the weight of the contrastive loss is set to $\alpha=1$. We split the fMRI sequence and corresponding text into multiple 20-second windows, with zero overlap. For the prompt length, we use $k=30$. For the mapping network, the BERT representation will be first mapped to a 512-dimensional vector before passing forward to the transformer mapping network. We use an 8-layer transformer here, with 8 attention heads in each layer. The fMRI encoder has the same architecture as the mapping network, except for a linear layer for the input. All the codes are implemented using PyTorch \cite{paszke2019pytorch}, and trained on an Nvidia A-100 GPU with the AdamW optimizer \cite{loshchilov2017decoupled}. The batch size is set to 32.

\subsection{Baseline and Evaluation Metrics}
We compare our method to Tang et al. \cite{tang2023semantic}, which is the state-of-the-art method in the dataset we used.  In their method, they employ a neuron encoding structure, that utilizes the LLM to generate proposal words and encode these proposals into fMRI to identify the most matching word. For a fair comparison, we employ the same story ("Where There’s Smoke") from the dataset as the test set and divide the entire story into 20-second windows, calculate evaluation metrics within each window, and average the scores of all windows to obtain the metric score for the entire test set. 

Similar to Tang et al. \cite{tang2023semantic}, we utilize identical language similarity metrics to evaluate our method in several aspects. BLEU \cite{papineni2002bleu} measures the number of individual translated segments that appear in the ground-truth text. METEOR \cite{denkowski2014meteor} computes the harmonic mean of unigram precision and recall. And BERTScore \cite{zhang2019bertscore} computes a similarity score for each token in the candidate sentence with each token in the reference sentence using contextual embeddings.

\subsection{Evaluation the Text Prompt}\label{sec: exp-T2T}
Since we treat the text prompt as the optimal prompt for decoding in our method, we first evaluate the performance of the text prompt in our work. We consider four settings in this part. For the first two settings, we use the word rate model in the inference and fine-tune or fix the parameters of GPT-2 in the training. For the last two settings, we use the alignment tokens and also compare two options for fine-tuning or fixing the parameters of GPT-2.

\begin{table}[h]
\caption{Text prompt performance. WR refers to the use of a word rate model in the inference. We mark the result that GPT-2 is not fine-tuned with ($\not$F).}
{\begin{tabular}{m{0.5cm} m{0.45cm}m{0.45cm}m{0.45cm} m{0.45cm}m{0.45cm}m{0.45cm} m{0.45cm}m{0.45cm}m{0.45cm}}
\hline
\multirow{2}{*}{}  & \multicolumn{3}{c}{\textbf{BLEU-1}$\uparrow$} & \multicolumn{3}{c}{\textbf{METEOR}$\uparrow$} & \multicolumn{3}{c}{\textbf{BERTScore}$\uparrow$} \\ \cline{2-10} 
                  & UTS1       & UTS2       & UTS3      & UTS1       & UTS2       & UTS3      & UTS1        & UTS2       & UTS3       \\ \hline
WR($\not$F)     & 19.68  &   20.85    &  18.62   &  14.14   &   14.66    &   12.66
  &   81.92  &   82.05  &   81.63
  \\ 
WR               & 22.96  &   24.21    &  24.10   &  16.92   &   16.99    &   17.60
  &   82.42  &   82.78  &   82.59
  \\ \hline
Our($\not$F)         & 21.89  &  20.44    &  21.87   &  20.32   &   20.40    &   20.82
  &   83.25  &   83.28  &   83.43
  \\ 
Our              & \textbf{26.21}  &   \textbf{26.13}    &  \textbf{25.54}   &  \textbf{26.27}   &   \textbf{25.97}    &   \textbf{25.49}
  &   \textbf{84.32}  &   \textbf{84.51}  &   \textbf{84.17}
  \\ \hline
\end{tabular}}
\label{tab: T2T-performance}
\end{table}

The experiment results are listed in Table \ref{tab: T2T-performance}. From the experimental results, we can find that the text prompt can be effectively decoded into text under various experimental settings. This result supports our further application of the text prompt in fMRI-to-text decoding. Also, we find that fine-tuning the GPT-2 in the training can always bring improvements in performance, regardless of the inference strategy we used. For the inference strategies, rather than using a word rate model to infer the text length, we find that adding alignment tokens in the ground-truth text in the training stage can effectively improve the decoding performance. Moreover, we find that fine-tuning the GPT-2 for the alignment tokens can bring a significant improvement in the results. Specifically, this setting can bring up to $5.69\%$ on BLEU-1, $5.95\%$ on METEOR, and $1.23\%$ on BERTScore among all the subjects.

\subsection{Evaluation of fMRI to Text Decoding}
In this part, we evaluate the decoding performance of our model by comparing it with the existing work \cite{tang2023semantic}. For the setting of our method, we add alignment tokens in the ground-truth text for the training. The GPT-2 is fine-tuned in the training, and the inference is stopped when the GPT-2 generates enough \$.

\begin{table}[h]
\caption{Compare our method with the existing work.}
{\begin{tabular}{m{0.5cm} m{0.45cm}m{0.45cm}m{0.45cm} m{0.45cm}m{0.45cm}m{0.45cm} m{0.45cm}m{0.45cm}m{0.45cm}}
\hline
\multirow{2}{*}{}  & \multicolumn{3}{c}{\textbf{BLEU-1}$\uparrow$} & \multicolumn{3}{c}{\textbf{METEOR}$\uparrow$} & \multicolumn{3}{c}{\textbf{BERTScore}$\uparrow$} \\ \cline{2-10} 
                 & UTS1       & UTS2       & UTS3      & UTS1       & UTS2       & UTS3      & UTS1        & UTS2       & UTS3       \\ \hline
Tang      & \textbf{23.31}     & \textbf{24.26}     & \textbf{24.70}    & 16.21     & 16.77     & 17.03    & 80.77      & 81.04      & 81.16     \\ \hline
BP-GPT               & 21.59  &   21.11    &  21.13   &  \textbf{20.82}   &   \textbf{19.76}    &   \textbf{20.34}
  &   \textbf{83.20}  &   \textbf{83.22}  &   \textbf{83.32}
  \\ \hline
\end{tabular}}
\label{tab: performance}
\end{table}

As is shown in Table \ref{tab: performance}, our method achieves comparable or even better performance. Specifically, on the METEOR, our method can achieve an improvement ranging from $2.99\%$ to $4.61\%$ on all the subjects. For the BERTScore, we also achieve an improvement ranging from $1.87\%$ to $2.43\%$ on all the subjects. 

\subsection{Ablation Study}
\begin{table}[h]
\caption{Ablation study: contrastive learning. We specify the method without contrastive learning with ($\not$C).}
{\begin{tabular}{m{0.5cm} m{0.45cm}m{0.45cm}m{0.45cm} m{0.45cm}m{0.45cm}m{0.45cm} m{0.45cm}m{0.45cm}m{0.45cm}}
\hline
\multirow{2}{*}{}  & \multicolumn{3}{c}{\textbf{BLEU-1}$\uparrow$} & \multicolumn{3}{c}{\textbf{METEOR}$\uparrow$} & \multicolumn{3}{c}{\textbf{BERTScore}$\uparrow$} \\ \cline{2-10} 
                  & UTS1       & UTS2       & UTS3      & UTS1       & UTS2       & UTS3      & UTS1        & UTS2       & UTS3       \\ \hline
WR($\not$C)         & 18.55  &   18.51    &  19.79   &  13.72  &   13.58    &   14.42  &   81.34  &   81.31 &   81.72
  \\ \hline               
WR         & 20.52  &   19.44    &  20.17   &  14.51  &   14.76    &   14.97  &   81.92  &   81.64  &   81.98
  \\ \hline
BP-GPT($\not$C)       & 20.27  &   20.41    &  20.43   &  19.43  &   19.58   &   19.46  &   82.90  &   82.78  &   82.81
  \\ \hline
BP-GPT               & \textbf{21.59}  &   \textbf{21.11}    &  \textbf{21.13}   &  \textbf{20.82}   &   \textbf{19.76}    &   \textbf{20.34}
  &   \textbf{83.20}  &   \textbf{83.22}  &   \textbf{83.32}
  \\ \hline
\end{tabular}}
\label{tab: Abalation-contrastive}
\end{table}

\subsubsection{Contrastive Learning}\label{sec: contras} To demonstrate the effectiveness of contrastive learning, we compare the performance of our method with or without contrastive learning. We include the experiment results on both inference strategies since the challenge of the significant modal difference between the fMRI and text exists in both settings. We aim to explore whether aligning the fMRI prompt with the text prompt can bring performance improvements in various inference strategies. As the target of the contrastive learning, the text prompt is extracted using the text-to-text baseline that has the same inference strategy as the fMRI-to-text model. We would like to refer to Section \ref{sec: exp-T2T} for more details.

We report the experiment results in Table \ref{tab: Abalation-contrastive}. By comparing Table \ref{tab: Abalation-contrastive} with Table \ref{tab: T2T-performance}, we can find that there is a significant gap between the performance of fMRI-to-text decoding and the performance of the text-to-text decoding, indicating the modal differences between text and fMRI impact the decoding performance seriously. Also, through the results, we find that aligning the fMRI prompt with the text prompt always brings performance improvements, no matter what inference strategies have been chosen. Specifically, when using a word rate model at the inference stage, aligning the fMRI prompt to the text prompt can bring an improvement up to $1.97\%$ on BLEU-1, $1.18\%$ on METEOR, and $0.58\%$ on BERTScore. While choosing the alignment token for inference, contrastive learning can bring an improvement up to $1.32\%$ on BLEU-1, $1.39\%$ on METEOR, and $0.44\%$ on BERTScore. This result proves that our approach of aligning fMRI prompts with text prompts is feasible and effective.

\subsubsection{Inference Strategy} Due to the characteristics of decoding tasks in auditory decoding scenarios, the choice of inference strategy has become particularly important. We further conduct an ablation study on it. Specifically, we compared four experimental settings, including the performance of two inference schemes with a fine-tuned or not fine-tuned GPT-2. The results are reported in Table \ref{tab: Abalation-inference}. Also, same as Section \ref{sec: contras}, the aligning target is the corresponding text prompt under the same inference strategy.

\begin{table}[h]
\caption{Abalation study: inference strategy.}
{\begin{tabular}{m{0.5cm} m{0.45cm}m{0.45cm}m{0.45cm} m{0.45cm}m{0.45cm}m{0.45cm} m{0.45cm}m{0.45cm}m{0.45cm}}
\hline
\multirow{2}{*}{}  & \multicolumn{3}{c}{\textbf{BLEU-1}$\uparrow$} & \multicolumn{3}{c}{\textbf{METEOR}$\uparrow$} & \multicolumn{3}{c}{\textbf{BERTScore}$\uparrow$} \\ \cline{2-10} 
                  & UTS1       & UTS2       & UTS3      & UTS1       & UTS2       & UTS3      & UTS1        & UTS2       & UTS3       \\ \hline
WR($\not$F)         & 19.80  &   19.36    &  19.97   &  13.43  &   14.09    &   13.90  &   81.57  &   81.45  &   81.29
  \\ \hline
WR         & 20.52  &   19.44    &  20.17   &  14.51  &   14.76   &   14.97  &   81.92  &   81.64  &   81.98
  \\ \hline
BP-GPT($\not$F)     & 20.83  &   19.49    &  20.65   &  20.60  &   19.24    &   20.22  &   83.18  &   82.95  &   82.98
  \\ \hline
BP-GPT              & \textbf{21.59}  &   \textbf{21.11}    &  \textbf{21.13}   &  \textbf{20.82}   &   \textbf{19.76}    &   \textbf{20.34}
  &   \textbf{83.20}  &   \textbf{83.22}  &   \textbf{83.32}
  \\ \hline
\end{tabular}}
\label{tab: Abalation-inference}
\end{table}

As is shown in Table \ref{tab: Abalation-inference}, we find that using alignment tokens to indicate the end of decoding can always bring a performance improvement. Also, fine-tuning the GPT-2 can bring more improvement with the alignment tokens. We believe that fine-tuning the GPT-2 in training can make the parameters of both the fMRI encoder and GPT-2 adapt to the alignment token. However, if fine-tuning is not performed, only the fMRI encoder will learn to adjust the fMRI prompt to enable GPT-2 to output \$ at the end of each text fragment that corresponds to a fMRI TR, bringing a negative influence on the performance.

\section{Conclusion}
In this work, we propose a decoding method capable of extracting text from fMRI signals within the auditory neural decoding scenario. The basic idea of our method involves employing an fMRI-prompted Large Language Model (LLM) for decoding. Furthermore, with the ongoing advancement of LLMs, our method remains readily compatible with updated and superior LLMs, facilitating performance improvements effortlessly.

\section{acknowledgments}
This work was supported in part by the Scientific and Technological Innovation (STI) 2030–Major Projects under Grant 2021ZD0201503; in part by Beijing Natural Science Foundation under Grant L243016; and  in part by the National Natural Science Foundation of China under Grant 62206284.

\section{Compliance with Ethical Standards}
This research study was conducted retrospectively using human subject data made available in open access by openneuro (https://openneuro.org/datasets/ds003020/versions/2.2.0). Ethical approval was not required as confirmed by the license attached with the open access data.


\bibliographystyle{IEEEtran}
\bibliography{BP-GPT}


\end{document}